\documentclass{elsarticle}
\newcommand{\text}[1]{\mbox{\scriptsize{#1}}}

\usepackage{lineno,hyperref}
\usepackage{epsfig,amsmath,amssymb,natbib}
\usepackage{color}
\modulolinenumbers[5]

\journal{International Journal of  Multiphase Flow}

\newcommand{\red}[1]{\textcolor{black}{#1}}
\newcommand{\blue}[1]{\textcolor{black}{#1}}

\biboptions{authoryear}

\begin{document}

\begin{frontmatter}

\title{A novel microfluidic method to produce monodisperse micrometer bubbles}

\author[label1]{A. Rubio}
\author[label1]{S. Rodr\'{\i}guez-Aparicio}
\author[label1]{J. M. Montanero\corref{cor1}}
\author[label1]{M. G. Cabezas}
\address[label1]{Departamento\ de Ingenier\'{\i}a Mec\'anica, Energ\'etica y de los Materiales and\\ 
Instituto de Computaci\'on Cient\'{\i}fica Avanzada (ICCAEx),\\ Universidad de Extremadura, E-06006 Badajoz, Spain}
\cortext[cor1]{Corresponding author: J. M. Montanero (jmm@unex.es)}

\begin{abstract}
We present a novel microfluidic method to produce quasi-monodisperse bubbles with diameters from tens to very few microns. A gaseous rivulet flows over the shallow groove printed on a T-junction exit channel. The triple contact line delimiting the rivulet is pinned to the groove edges. The rivulet breaks up into bubbles much smaller than the exit channel. When operating under adequate conditions, the flow transitions toward a singular mode where the rivulet remains quasi-static and emits bubbles smaller than the groove width. This allows the production of bubbles with diameters in the 3-5 $\mu$m range, which is preferable for relevant therapeutical applications.
\end{abstract}


\end{frontmatter}


\section{Introduction}

The production of monodisperse collections of microbubbles is essential in fields such as medicine \citep{SE08}, pharmacology \citep{FPB07}, material science \citep{SP99}, and the food industry \citep{ZA08}. In particular, microbubbles are the most effective contrast agent for medical ultrasound imaging \citep{FSLT20}. They are used in therapeutic applications, including sonoporation, tumor ablation, and sonothrombolysis, and can be carriers of gas, genes, and oxygen \citep{UPCLMZ04}. Bubbles with diameters below 8 $\mu$m and low polydispersity indexes must be produced at sufficiently large rates for medical applications. 

Two- and three-dimensional co-flow, cross-flow, flow focusing, and T-junction microfluidic devices have been widely used to produce monodisperse collections of microbubbles \citep{SSA04,CA07,A16}. In these cases, the sizes of the fluid passages are similar to or even smaller than the size of the bubble, which constitutes a serious drawback in terms of the high pressures demanded, the limited production rates, and the device clogging. 

In an axisymmetric flow-focusing device \citep{GG01,G04b,VAMHG14}, a coflowing liquid stream crosses an orifice located in front of the gas source. The viscous and pressure forces exerted by that stream collaborate to stretch a gaseous tapering meniscus attached to the feeding capillary, significantly reducing the bubble size at high production rates. However, the bubbles produced with this method are at least tens of microns in diameter when the focusing liquid is water, even if its surface tension is lowered by adding a surfactant. The planar version of the flow-focusing device \citep{ABS03,GGDWKS04,HTLDL07,DHRMV08} has been broadly used to produce monodisperse bubbles for therapeutic applications. In this case, the bubble formation is geometrically controlled, which increases the degree of monodispersity. However, the bubble is commensurate with the channel size. 

\citet{CHLG11} produced bubbles with sizes below 10 $\mu$m by applying the flow focusing principle in a 3D device with quadrangular channels 50 $\mu$m in width. The emitted gaseous thread was stabilized by forcing the pinning of the triple contact line to the boundary between a centered hydrophobic strip and the surrounding hydrophilic surface of one of the walls of the discharge channel. However, the gas ligament was not straight, reducing the monodispersity degree and bubbling frequency \citep{CRG16}. In addition, the device could be used only for 24 hours to ensure surface hydrophobicity \citep{CHLG11}. These limitations were overcome by the device manufactured by \citet{CRG16}, in which the hydrophobic surface was substituted by a groove of around 7.4 $\mu$m in width and 5 $\mu$m in depth. The minimum bubble diameter obtained with this device was around 9 $\mu$m, exceeding the maximum value for therapeutical applications. \citet{HGM13} proposed a method in which the gas adhered to a hydrophobic strip printed on the exit channel of a T-junction. The method was studied numerically; it has not been implemented experimentally.

We propose a method to produce microbubbles considering the ideas of \citet{CHLG11}, \citet{HGM13}, and \citet{CRG16}. In this method, the two fluids meet in a T-junction. The liquid current forces a micrometer gaseous rivulet to slip over the bottom of the discharge channel, in which a narrow groove was printed to pin the triple contact line. The groove depth was very small to minimize gas flow under the channel surface. As explained in Sec.\ \ref{sec2}, microbubbles are produced from the rivulet breakup following an inertio-capillary mechanism. 

Our method verifies two essential conditions commonly demanded in medical applications: (i) bubbles are smaller than 8 $\mu$m in diameter, and (ii) the polydispersity index (the ratio of the standard deviation to the mean value) is smaller than 10\%. Besides, all the device passages are much bigger than the bubbles produced, allowing the device to work safely without clogging and with relatively small applied pressures. 

\section{The proposed method. Bubble ejection modes}
\label{sec2}

\begin{figure}[hbt]
\begin{center}
\resizebox{0.9\textwidth}{!}{\includegraphics{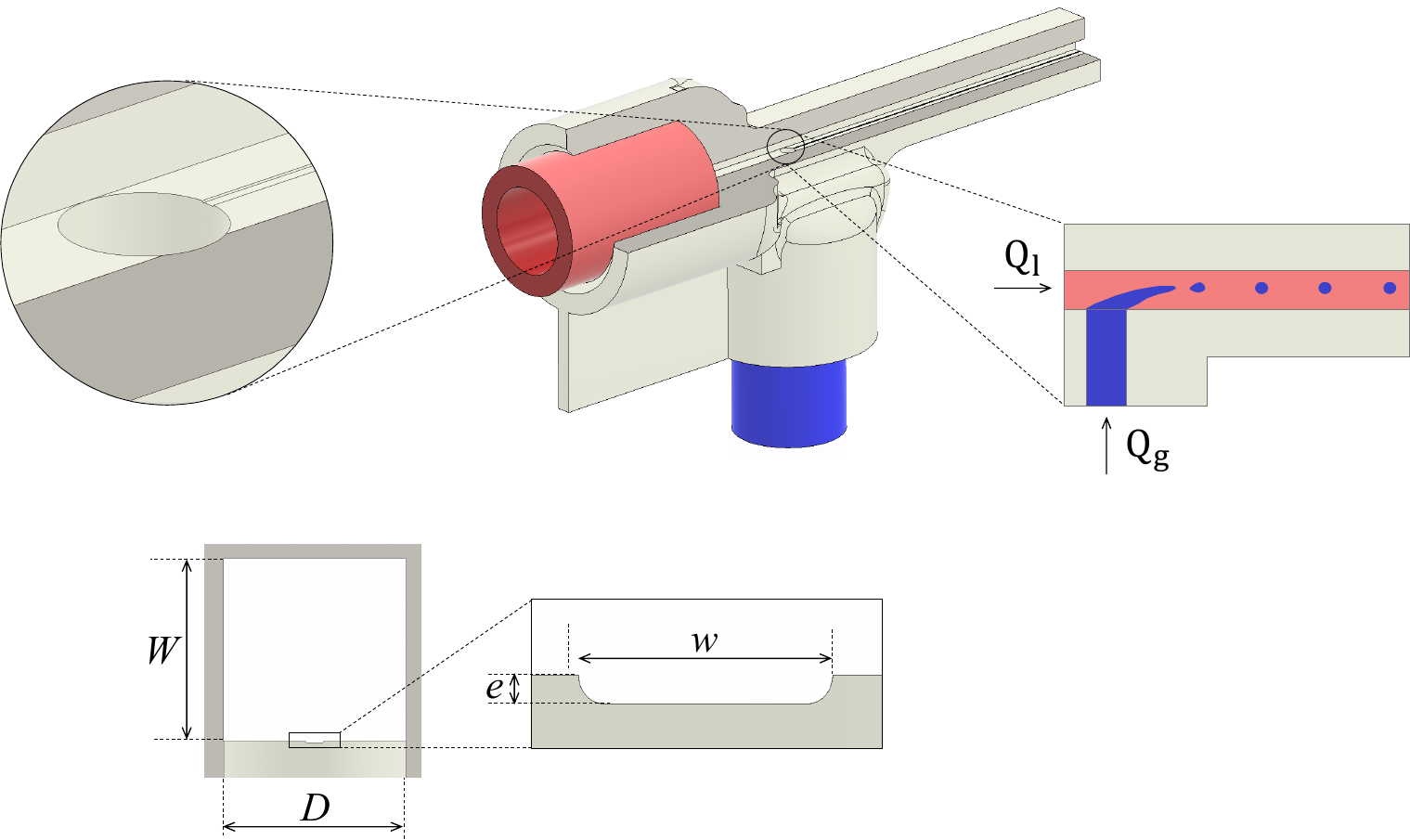}}
\end{center}
\caption{Microfluidic device used in the experiments.}  
\label{device}
\end{figure}

Figure \ref{device} shows an image of the microfluidic device used to produce microbubbles. Water is injected at a constant flow rate $Q_l$ through the horizontal red tube connected to the quadrangular channel of width $W$. Air is injected at a constant flow rate $Q_g$ across the blue tube of diameter $D$, also coupled to the quadrangular channel through a T-junction. A shallow groove of width $w$ and depth $e$ is printed on the discharge channel bottom.  

\begin{figure}[hbt]
\begin{center}
\resizebox{0.65\textwidth}{!}{\includegraphics{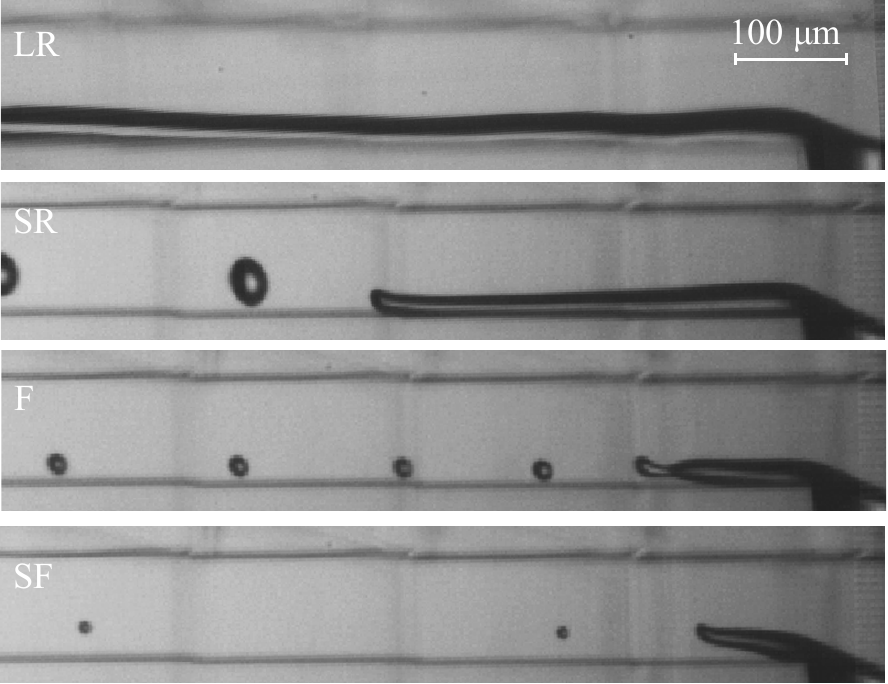}}
\end{center}
\caption{Images of the long-rivuletting (LR), short-rivuletting (SR), focusing (F), and singular focusing (SF) modes obtained for $Q_g=4.41$ ml/h, $0.64$ ml/h, $0.09$ ml/h, and $0.001$ ml/h, respectively. In the three cases, $Q_l=90$ ml/h. The videos are shown in the Supplemental Material.}  
\label{modes}
\end{figure}

Figure \ref{modes} shows the flow modes found in our experiments as the gaseous flow rate $Q_g$ decreases for a fixed liquid flow rate $Q_l$. When the gas enters the T-junction, it flows toward the groove driven by the liquid stream. The triple contact lines pins to the groove edges, and a long rivulet moves over the groove. The rivulet cross-section area is practically constant, as occurs in the classical jetting regime. The flow is convectively unstable, which means that capillary waves are convected downstream, allowing the formation of a long, stable gaseous thread. The gas flow rate controls the volume of the rivulet, whose end breaks up into a quasi-monodisperse collection of bubbles due to the capillary instability \citep{HMMG15}. The size of the bubble is commensurate with that of the rivulet, analogously to what occurs in the classical liquid jetting mode. Hereafter, we will refer to this flow as the ``long-rivuletting" (LR) mode. 

At a given gas flow rate, the rivulet described above sharply shortens. The gas accelerates (the cross-section area decreases) along the gaseous thread. As in the LR mode, the rivulet end breaks up into quasi-monodisperse bubbles with sizes that are commensurate with that of the rivulet. We will refer to this flow as the ``short-rivuletting" (SR) mode (Fig.\ \ref{modes}). In both the LR and SR modes, the triple contact line in the rivulet front is not pinned but oscillates during the bubble detachment. The difference between the LR and SR modes lies in the character of the instability growing in the rivulet: the rivulet is convectively (absolutely) unstable \citep{HM90a} in the LR (SR) mode. A similar distinction can be made, for instance, in the liquid-liquid coflow configuration \citep{M24}.

Interestingly, at a critical gas flow rate, $Q_{gc}$, the flow transitions to another regime in which the rivulet tip detaches from the solid surface and ejects tiny bubbles (Fig.\ \ref{modes}). Unlike in the LR and SR modes, the front triple contact line is pinned, favoring the energy focusing. We will call this behavior the ``focusing" (F) mode. 

We decreased the gaseous flow rate slowly and in very small steps to produce the F mode for $Q_g<Q_{gc}$. At a certain point, the flow autonomously adopted a singular version of this mode characterized by a much smaller gas flow rate (Fig.\ \ref{modes}). The rivulet becomes a quasi-static gaseous thread. The liquid current drives the gas to the thread tip. Pressure is built up there due to this hydrodynamic focusing effect, allowing the formation of bubbles with diameters even smaller than the groove width. Hereafter, we will refer to this flow as the ``singular-focusing" (SF) mode. The transition from the F to the SF mode seems to be linked to the gas flow rate fluctuations caused by the bubble ejection. The relative magnitude of these fluctuations becomes significant in this mode because $Q_g$ takes very small values, and the capillary pressure considerably fluctuates during the ejection of bubbles due to their small size. 

The SF mode resembles the microbubbling regime of flow focusing \citep{GG01}, where a quasi-static gaseous meniscus emits tiny bubbles from its tip. As shown in Sec.\ \ref{sec4}, this mode produces quasi-monodisperse collections of bubbles with diameters well below 8 $\mu$m, the threshold for medical applications. It is robust (the flow remains stable indefinitely) and highly reproducible. When $Q_g$ is decreased below the SF mode value, the gaseous stream does not continuously enter the liquid channel.  

The critical role played by the groove in the flow behavior described above must be pointed out. \blue{The groove fixes the size of the rivulet below that of the channel. The triple contact line cannot move outward, and the gas remains confined in a small region of the channel}. Without that groove, one obtains the classical bubbling and slugging regimes, which give rise to bubbles that are commensurate with or are even larger than the discharge channel size. The groove in our experiments is hydrophilic and much shallower than in previous experiments \citep{CRG16}. This is key to lowering the bubble size down to 3 $\mu$m \blue{because it allows us to reduce the gas volume transported by the rivulet}. The sequence of modes described above occurs only within a relatively narrow interval of the liquid flow rate. That interval depends on the microfluidic device's characteristic lengths $w$ and $W$. 

\section{Experimental method}
\label{sec3}

The microfluidic device (Fig.\ \ref{device}) was printed using Nanoscribe Photonic Professional GT2 with the Dip-in Laser Lithography (DiLL) configuration. The 25$\times$ objective was dipped into the IP-S resin droplet deposited on an ITO-coated glass substrate. We chose the solid writing strategy. The typical slicing and hatching distances were $1$ $\mu$m and 0.5 $\mu$m. The part was developed in $\sim$25 ml of propylene glycol monomethyl ether acetate (PG-MEA) for 24 h and then cleaned in isopropanol for 2 h. Then, unexposed resin inside the shell was cured for 60 min inside the UV Curing Chamber (XYZprinting). The experiments were conducted with three devices: ($W=100$ $\mu$m, $D=100$ $\mu$m, $w=10$ $\mu$m), ($W=50$ $\mu$m, $D=50$ $\mu$m, $w=5$ $\mu$m),  and ($W=100$ $\mu$m, $D=100$ $\mu$m, $w=5$ $\mu$m). The groove depth was $e=1$ $\mu$m in the three cases.

\begin{figure}[tbp]
\begin{center}
\resizebox{0.7\textwidth}{!}{\includegraphics{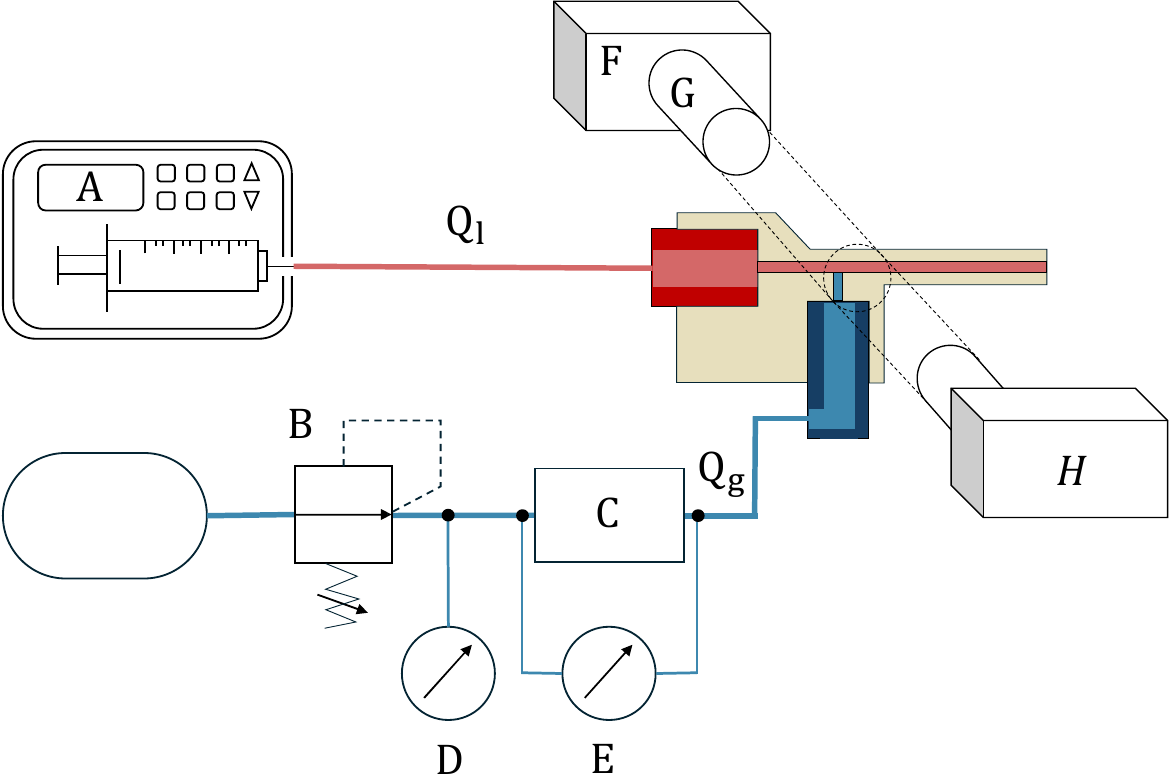}}   
\end{center}
\caption{Experimental setup: syringe pump (A), pressure regulating valve (B), hydraulic resistance (C), high-precision manometer (D), differential pressure gauge (E), camera (F), and optical lenses (G), and optical fiber connected to a light source (H). The injected liquid and gas flow rates are $Q_l$ and $Q_g$, respectively.}
\label{setup}
\end{figure}

Figure \ref{setup} shows the experimental apparatus used in this work. The liquid was injected with a syringe pump (A). The gas pressure was controlled with a high-precision pressure regulating valve (B). Then, the air entered into a hydraulic resistance (C) 155 cm in length and 160 $\mu$m in inner diameter, which allowed us to fix the gas flow rate and eliminate any mechanical perturbation originating at the pressure valve. The hydraulic resistance supplied the air stream to the microfluidic device. Both the gauge pressure at the tank exit and the pressure drop in the hydraulic resistance were measured with a high-precision manometer (D) and a differential pressure gauge (E), respectively. 

Digital images of the rivulet were acquired at up to 20\,000 frames per second with an exposure time of 367 ns using an ultra-high-speed CMOS camera ({\sc Photron, FASTCAM SA5}) (F). The camera was equipped with a set of optical lenses (G), which consisted of a 10$\times$ magnification zoom-objective (OPTEM HR) and a system of lenses (OPTEM 70 XL) with variable magnification from 1.5$\times$ to 5.25$\times$. The magnification obtained was approximately from 0.38 to 1.33 $\mu$m/pixel. The fluid configuration was illuminated from the back side by cool white light provided by an optical fiber connected to a light source (H). All these elements were mounted on an optical table.

The gas used in the experiments was air ($\rho_g=1.22$ kg/m$^3$, $\mu_g=18.5$ $\mu$Ps$\cdot$s), while the liquids were water ($\rho_l=998$ kg/m$^3$, $\mu_l=0.89$ mPa$\cdot$s) and distilled water with Tween 80 at the concentration 2\% (w/v) \citep{CHLG11}. The density and viscosity of the Tween 80 aqueous solution were practically the same as those of water, while the surface tension of the gas-liquid interface decreased from $\gamma=72$ mN/m to 39 mN/m. 

The advancing $\theta_a$ and receding $\theta_r$ contact angles of the working liquids on the device surface were measured with the sessile drop method \citep{KHIR13}. The values for water were $\theta_r=49^{\circ}$ and $\theta_a=80^{\circ}$, while the values for the Tween 80 aqueous solution were $\theta_r=6^{\circ}$ and $\theta_a=47^{\circ}$.

The experimental procedure was as follows. The device was carefully cleaned with isopropanol. Then, we fixed the liquid flow rate $Q_l$ and the gas flow rate $Q_g$. The gas flow rate was sufficiently high to produce the SR mode. Then, $Q_g$ was decreased while keeping $Q_l$ constant. A video of the bubble ejection was recorded for each pair of values $(Q_l,Q_g)$. The experiment was repeated for several values of $Q_l$. The bubble diameter $d_b$ and ejection frequency $f_b$ were measured from the images. The gas flow rate $Q_g$ at the discharge channel was obtained as $Q_g=f_b\, \pi d_b^3/6$. We did not analyze the LR mode because the rivulet did not break up into bubbles in many of the experimental realizations.

\section{Results}
\label{sec4}

\begin{figure}[hbt]
\begin{center}
\includegraphics[width=0.5\textwidth]{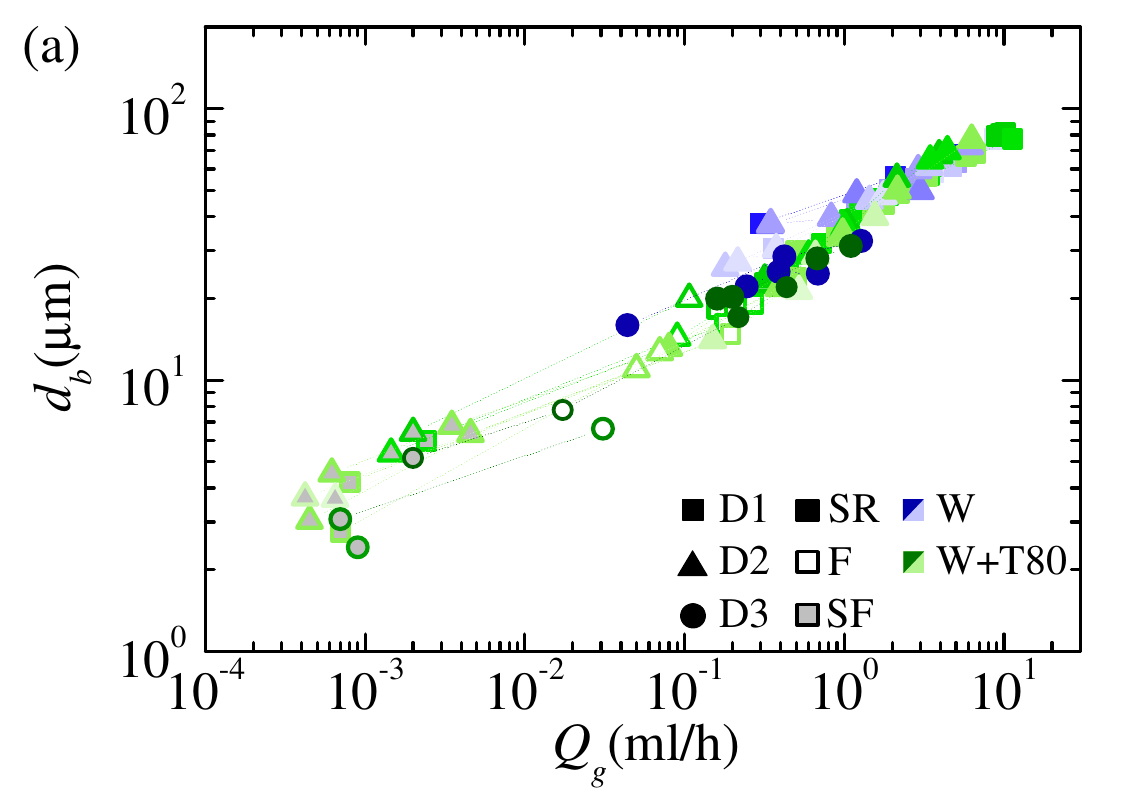}\includegraphics[width=0.5\textwidth]{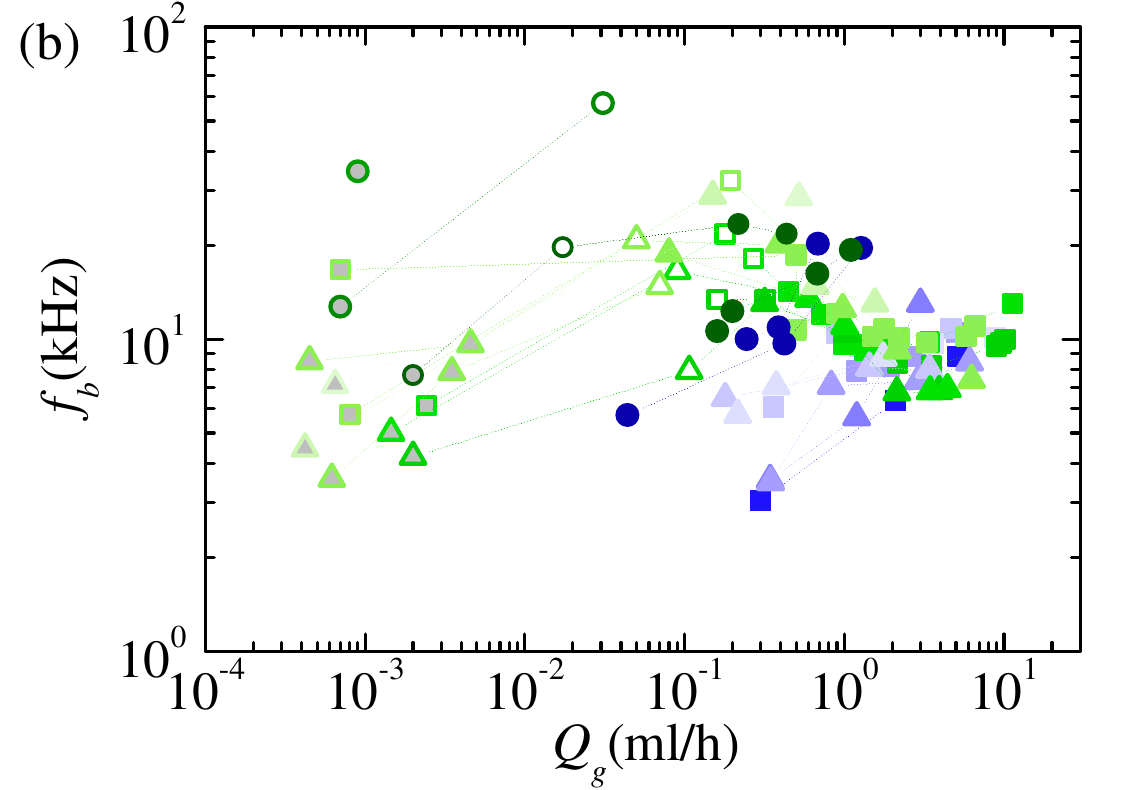}  
\end{center}
\caption{Bubble diameter $d_b$ (a) and production frequency $f_b$ (b) as a function of the gas flow rate $Q_g$ for $W=100$ $\mu$m and $w=10$ $\mu$m (D1), $W=100$ $\mu$m and $w=5$ $\mu$m (D2), and $W=50$ $\mu$m and $w=5$ $\mu$m (D3). The solid symbols correspond to the SR mode. The open and filled with grey symbols correspond to the F and SF modes, respectively. The blue and green symbols correspond to water (W) and water+Tween 80 (W+T80), respectively. Blue and green scales indicate different values of $Q_l$.}
\label{diameter2}
\end{figure}

\begin{figure}[hbt]
\begin{center}
\resizebox{0.5\textwidth}{!}{\hspace{-2.5cm}\includegraphics{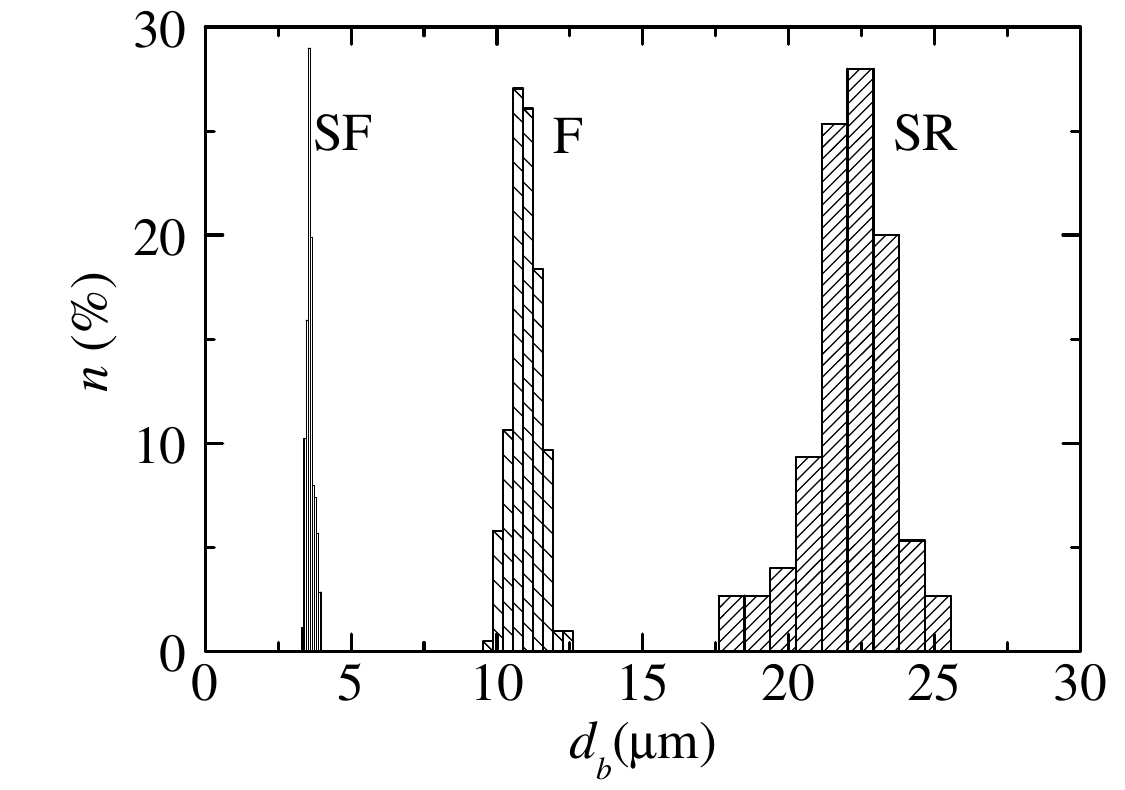}}
\end{center}
\caption{Normalized histogram of the bubble diameter for experimental realizations in the SR, F, and SF modes. The results were obtained for $W=100$ $\mu$m, $w=10$ $\mu$m, $Q_l=110$ ml/h, and $Q_g=0.39$ ml/h (SR mode), $0.052$ ml/h (F mode), and $0.00045$ ml/h (SF mode). The values of the PDI were around 7\%, 4\%, and 4\%, respectively. Here, the PDI value was calculated as $\sigma/\mu$, where $\sigma$ and $\mu$ are the standard deviation and mean value of the normal distribution fitted to the experimental data.} 
\label{pdi}
\end{figure}

Figure \ref{diameter2}a shows the bubble diameter $d_b$ for all the experimental realizations. The transition from the SR to the F mode allowed us to reduce the bubble diameter. The F mode remained stable for smaller values of $Q_g$ when $w$ was decreased for a fixed value of $W$. An extra stabilization effect was achieved by decreasing $W$ as well. Bubbles with diameters in the 6-15 $\mu$m range were produced in the F mode for $w=5$ $\mu$m. When the flow adopted the SF mode, the bubble diameter considerably decreased. Microbubbles with diameters smaller than 3 $\mu$m were produced at frequencies \red{$f_b=6Q_g/(\pi d_b^3)$} larger than 30 kHz (Fig.\ \ref{diameter2}b) with a high monodispersity degree (Fig.\ \ref{pdi}) in this mode. This diameter is two orders of magnitude smaller than the channel width. The SF mode was found in the three devices used in our experiments.

\red{The bubble diameter depends not only on the gas flow rate but also, to a lesser extent, on the rest of the governing parameters (channel size, groove width, liquid flow rate, and surface tension). This explains the scattering of the results in Fig.\ \ref{diameter2}a, where the bubble diameter is plotted versus the gas flow rate for all the experimental realizations. The scattering is around three times larger in Fig.\ \ref{diameter2}b because the bubble frequency is proportional to $d_b^{-3}$. Therefore, the scattering in Fig.\ \ref{diameter2} must not be attributed to instabilities or fluctuations during the bubble formation. These factors are quantified by the polydispersity index of the bubble diameter histogram obtained for a given experimental realization. As shown below, the values of this index were very small (around 5\%), which shows the robustness of the process.}

\begin{figure}[hbt]
\begin{center}
\resizebox{0.5\textwidth}{!}{\hspace{-2.5cm}\includegraphics{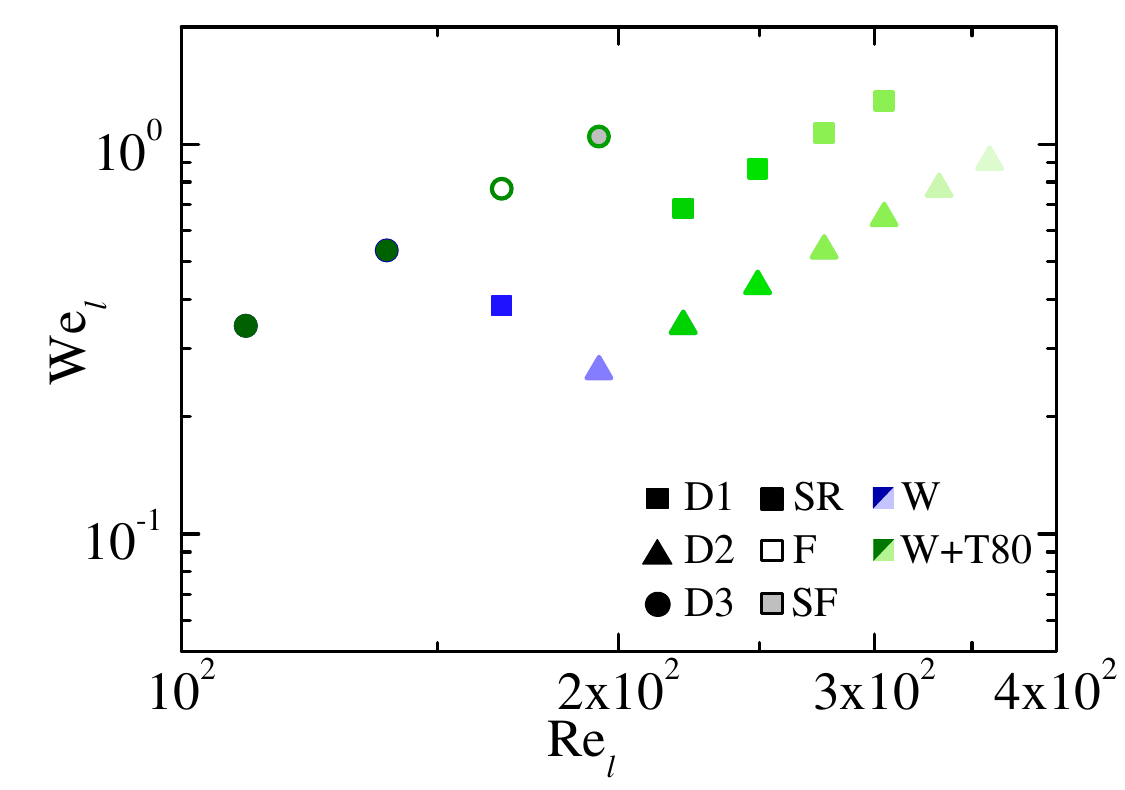}}
\end{center}
\caption{\blue{Reynolds Re$_l$ and Weber We$_l$ numbers in the experimental realizations.}} 
\label{ReWe}
\end{figure}

For a fixed geometry, the parameters involved in the problem are the width $W$ of the discharge channel, the gas density $\rho_g$ in the T-junction, the liquid density $\rho_l$, the gas viscosity $\mu_g$, the liquid viscosity $\mu_l$, the surface tension $\gamma$, the gas flow rate $Q_g$, and the liquid flow rate $Q_l$. Five dimensionless numbers can be formed with these parameters: the gas and viscosity ratios, $\rho_g/\rho_l$ and $\mu_g/\mu_l$, the Reynolds number Re$_l=\rho_l Q_l/(W\mu_l)$, the Weber number We$_l=\rho_l Q_l^2 w/(W^4\gamma)$, and the flow rate ratio $Q_g/Q_l$. The gas and viscosity ratios were fixed in our experiments (except for the small variations of $\rho_g$ due to its dependency on the gas injection pressure). \blue{The values of the Reynolds number lie in the interval $110\lesssim \text{Re}_l\lesssim 360$ (Fig.\ \ref{ReWe}), implying that the liquid viscosity plays a secondary role. The Weber number takes values of order unity, indicating that liquid inertia $\rho_l Q_l^2/W^4$ is commensurate with the capillary pressure $\gamma/w$.}  

Our experimental results can be rationalized in terms of the scaling analysis proposed by \citet{CHLG11}. Assuming that the flow along the gas rivulet is developed, $Q_g$ can be calculated as
\begin{equation}
Q_g=-K\, \frac{d_r^4}{\mu_g}\, \frac{dp_g}{dx},
\end{equation}
where $K$ is a constant, $d_r$ is a rivulet effective diameter, $dp_g/dx$ is the gas pressure gradient in the channel direction $x$, and we have neglected the gas flow rate due to the Couette-type flow driven by the liquid current. 

Suppose the rivulet cross-section area is approximately constant. Therefore, the capillary pressure does not significantly change along the rivulet, implying that $dp_g/dx=dp_l/dx$, where $dp_l/dx=C \mu_l U_l/W^2$ is the liquid pressure gradient in the channel direction, $C$ is a constant, and $U_l=Q_l/W^2$ is the liquid mean velocity. Then, one obtains 
\begin{equation}
\label{e1}
\frac{d_r}{W}\propto \left(\frac{\mu_g}{\mu_l}\right)^{1/4} \left(\frac{Q_g}{Q_l}\right)^{1/4}.
\end{equation}
In the SR and F modes, the pressure variations produced by the bubble ejection are not expected to produce significant variations in the flow rate transported by the rivulet. Under this condition, the bubble ejection frequency scales as $f_b\sim U_l/d_r$ \citep{RSMG15}. Therefore, 
\begin{equation}
\label{e2}
\frac{Q_g}{Q_l}=\frac{\pi d_b^3}{6}f_b\, \frac{1}{U_l W^2}\propto \frac{d_b^3}{d_r W^2}\, .
\end{equation}
Equations (\ref{e1}) and (\ref{e2}) lead to \citep{CHLG11}
\begin{equation}
\label{sc}
\frac{d_b}{W}\propto \left(\frac{\mu_g}{\mu_l}\right)^{1/12} \left(\frac{Q_g}{Q_l}\right)^{5/12}\, .
\end{equation}
Equation (\ref{sc}) predicts that the flow rate ratio $Q_g/Q_l$ essentially controls the dimensionless bubble diameter. The viscosity ratio (the gas viscosity) plays a secondary role. The effect of the Reynolds number and Weber number (the surface tension) is negligible.

\citet{CRG16} assumed that $d_r$ remained practically constant in their experiments with a groove; i.e., $d_r$ did not significantly depend on the flow rate ratio $Q_g/Q_l$. In this case,
\begin{equation}
\label{sc2}
\frac{d_b}{W}\propto \left(\frac{Q_g}{Q_l}\right)^{1/3}\, .
\end{equation}
In the experiments of \citet{CRG16}, the rivulet effective diameter $d_r$ was essentially determined by the groove width $w$. The groove in those experiments was much deeper than in our devices. For this reason, the condition $d_r\simeq \text{const.}$ is not expected to hold in all our experimental realizations.

\begin{figure}[hbt]
\begin{center}
\includegraphics[width=0.5\textwidth]{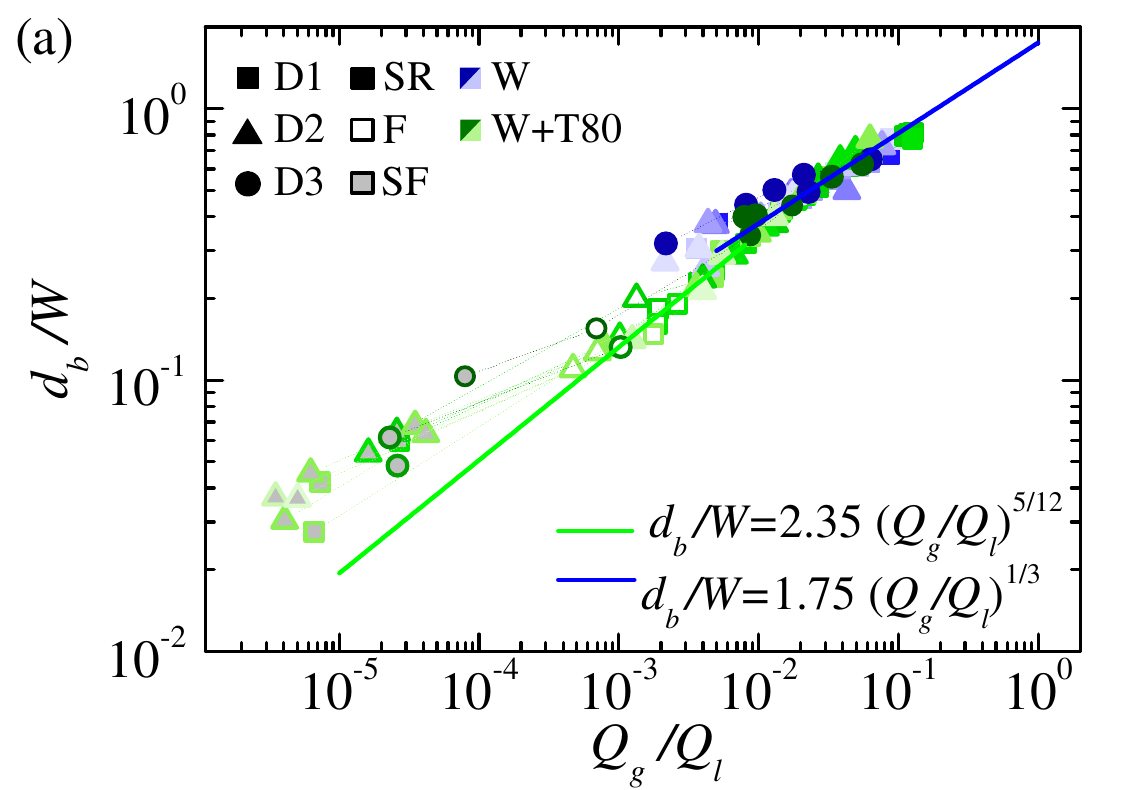}\includegraphics[width=0.5\textwidth]{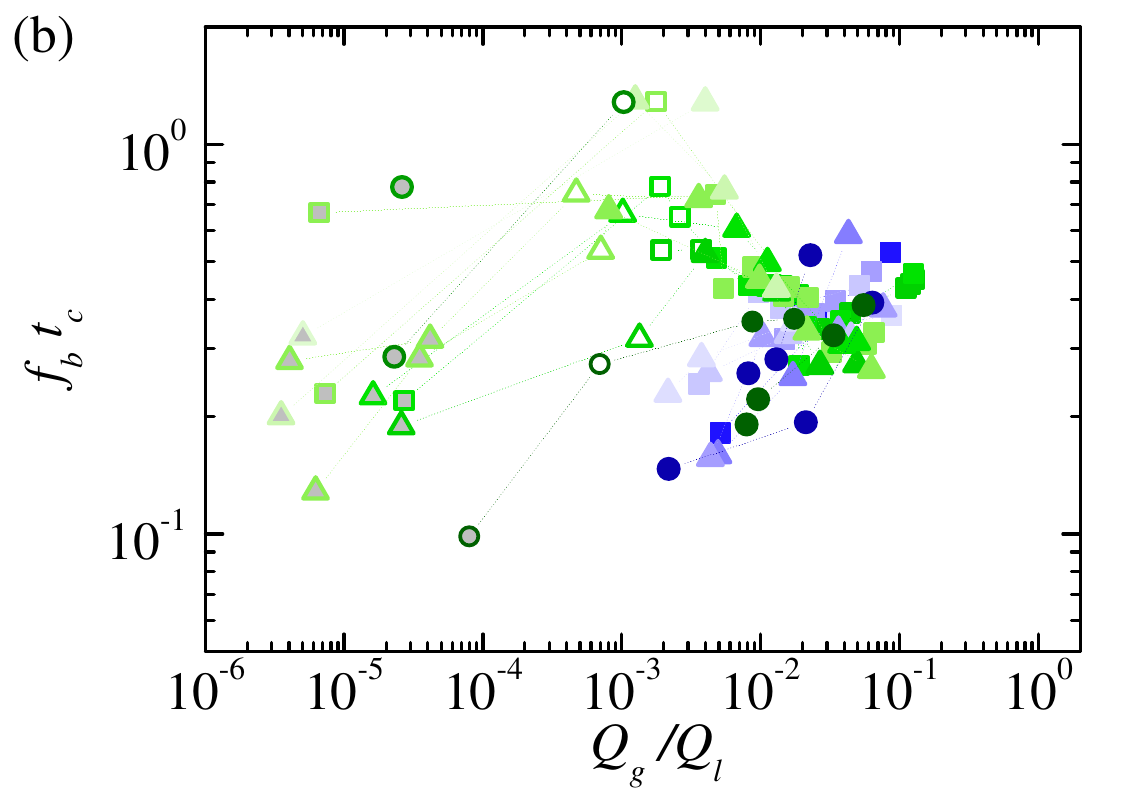}
\end{center}
\caption{Bubble diameter $d_b/W$ (a) and production frequency $f_b\, t_c$ ($t_c=W/U_l$) (b) as a function of the flow rate ratio $Q_g/Q_l$ for $W=100$ $\mu$m and $w=10$ $\mu$m (D1), $W=100$ $\mu$m and $w=5$ $\mu$m (D2), and $W=50$ $\mu$m and $w=5$ $\mu$m (D3). The solid symbols correspond to the SR mode. The open and filled with grey symbols correspond to the F and SF modes, respectively. The blue and green symbols correspond to water (W) and water+Tween 80 (W+T80), respectively. Blue and green scales indicate different values of Re$_l$.} 
\label{diameter}
\end{figure}

Figure \ref{diameter}a shows the bubble diameter $d_b/W$ as a function of the flow rate ratio $Q_g/Q_l$ in our experiments. The figure shows results obtained for $w/W=0.05$ and 0.1, with and without surfactant. The experimental data for the SR mode satisfactorily agree with the scaling law (\ref{sc2}), which indicates that $d_r$ is approximately constant in those experiments. The scaling law (\ref{sc}) approximately holds in the F mode, which suggests that $d_r$ decreases with  $Q_g/Q_l$ in this mode [Eq.\ (\ref{e1})]. The bubble diameter for $Q_g/Q_l\lesssim 10^{-3}$ exceeds the prediction of the scaling law (\ref{sc}) because of the variations in the gas flow rate during the bubble ejection \citep{RSMG15}. As a result, the scaling law (\ref{sc2}) reasonably agrees with the experimental data over the whole range of $Q_g/Q_l$. The narrow intervals of $U_l$ and $d_r$ in our experiments explain the relatively small variations of the bubble ejection frequency (Fig.\ \ref{diameter}b). 

\begin{figure}[hbt]
\begin{center}
\resizebox{0.5\textwidth}{!}{\hspace{-2.5cm}\includegraphics{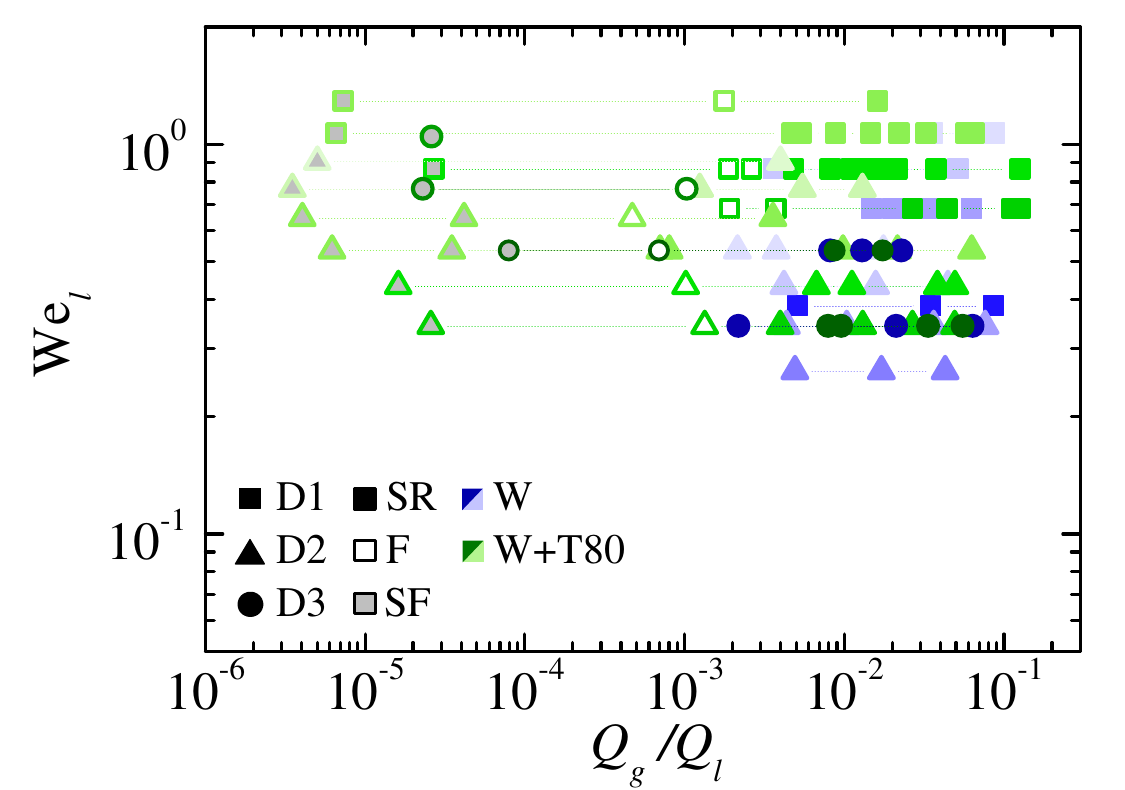}}
\end{center}
\caption{Stability map in the parameter plane ($Q_g/Q_l$,We$_l$) indicating the region where the different modes were obtained. The symbol shape indicate the device: $W=100$ $\mu$m and $w=10$ $\mu$m (D1), $W=100$ $\mu$m and $w=5$ $\mu$m (D2), and $W=50$ $\mu$m and $w=5$ $\mu$m (D3). The solid symbols correspond to the SR mode. The open and filled with grey symbols correspond to the F and SF modes, respectively. The blue and green symbols correspond to water and water+Tween 80 (W+T80), respectively. Blue and green scales are used to indicate different values of Re$_l$.} 
\label{map}
\end{figure}

\begin{figure}[hbt]
\begin{center}
\resizebox{0.5\textwidth}{!}{\hspace{-2.5cm}\includegraphics{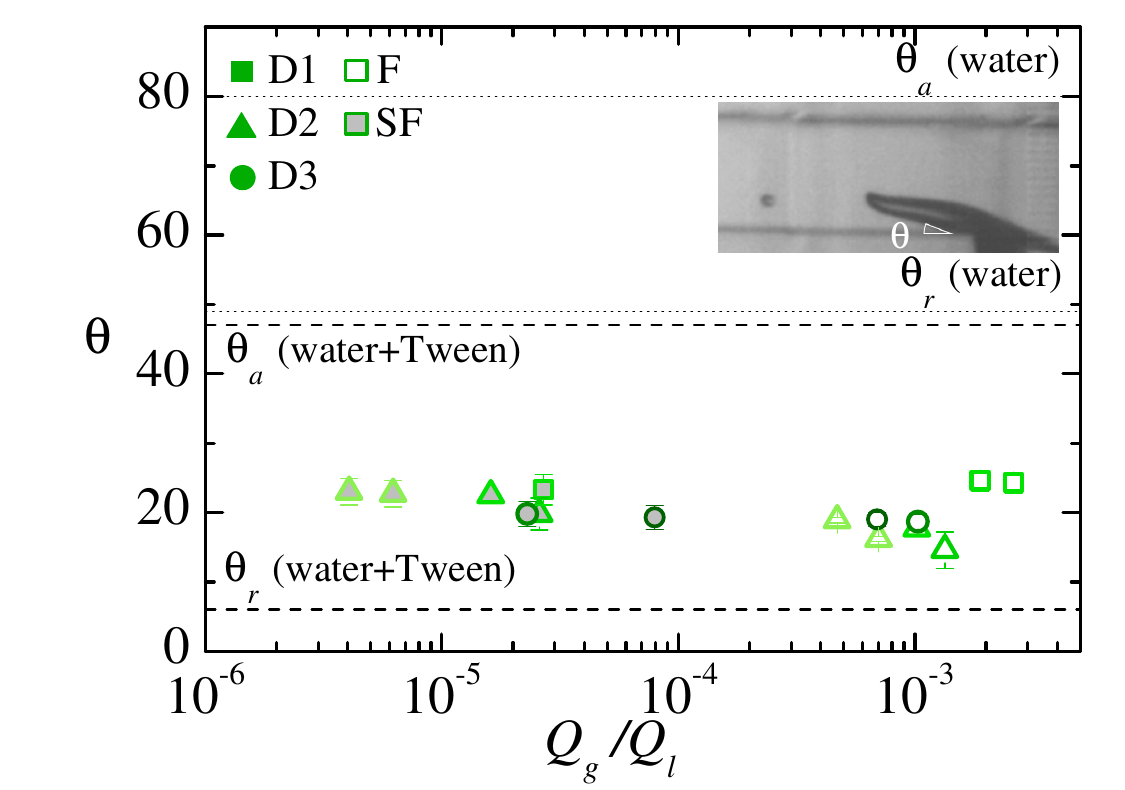}}
\end{center}
\caption{Contact angle $\theta$ estimated from the images of the F (open symbols) and SF (filled with grey symbols) modes. The dotted and dashed lines indicate the receding $\theta_r$ and advancing $\theta_a$ contact angles for water and water+Tween 80, respectively.} 
\label{con}
\end{figure}

The F and especially the SF mode allowed us to produce monodisperse bubbles that can be useful in medical applications. Figure \ref{map} shows the stability map in the parameter plane ($Q_g/Q_l$, We$_l$), indicating the region where those modes were obtained. \red{The results suggest that there are two requisites to obtain the F and SF modes: the flow rate ratio (the gas flow rate) and the receding contact angle must be low enough. When the gas flow rate is sufficiently small, momentum transferred by the liquid current builds up pressure in the rivulet tip. This pressure overcomes the capillary pressure. In addition, the surfactant lowers the receding contact angle so that the triple contact line in the rivulet front can pin to the channel surface ($\theta_r\leq \theta\leq \theta_a)$ (Fig.\ \ref{con}). Then, the rivulet tip detaches from the surface and ejects tiny bubbles.}

\section{Concluding remarks}

We have developed a T-junction-based device capable of producing quasi-monodisperse bubbles much smaller than any of its dimensions. The device relies on the pinning of the lateral three-phase lines delimiting a rivulet. This is achieved due to a groove in the exit channel, as in the flow focusing method proposed by \citet{CRG16}. However, when operating under adequate conditions, our device can produce bubbles with smaller diameters than the groove width. 

The new device can produce bubbles smaller than red blood cells (8 $\mu$m in diameter), making them suitable for ultrasound contrast agents. In particular, it is possible to produce quasi-monodisperse bubbles approximately 3-5 $\mu$m in diameter, which are preferable for some therapeutical applications because they are resonant to ultrasound frequencies used for therapy \citep{FSLT20}.

The major drawback of the proposed method is probably the relatively small bubble production frequency (of the order of tens of kHz), a characteristic inherent to the T-junction geometry. The flow-focusing (cross-flow) configuration may operate with larger liquid velocities (applied pressures) and smaller effective rivulet diameters, which may overcome this limitation. 

Two-photon polymerization allows one to optimize the flow geometry, leading to a new generation of microfluidic devices for microbubble production that satisfy the stringent conditions demanded in many applications.

\vspace{1cm}

This work was financially supported by the Spanish Ministry of Science, Innovation and Universities (grant no. PID2022-140951OB-C22/AEI/10.13039/501100011033/FEDER, UE). 


\end{document}